# Multi-criteria optimization and automated network restructuring to mitigate construction projects delays on-the-run

## - an extension of the Mitigation Controller (MitC) -

N.(Nina) Prins[1], O. (Omar) Kammouh[2], A.R.M. (Rogier) Wolfert[3]

**Abstract (250 words)**

**Purpose (limit 100 words)**

Construction project management requires dynamic mitigation control ensuring the project's timely completion by a best fit for common purpose strategy for all stakeholders. Current mitigation approaches are usually performed by an iterative Monte Carlo (MC) analysis focussing on lowest-cost strategies which do not include (1) the project manager's goal-oriented behaviour, (2) automated network restructuring potential, and (3) multi-dimensional optimization criteria for best fitting mitigation strategies-criteria. Therefore, the development statement within this paper is to design a method and implementation tool that properly dissolves all the aforementioned shortcomings ensuring the project's completion date by finding the most effective and efficient mitigation strategy.

**Design/methodology/approach (limit 100 words)**

To fulfill the purpose of this paper, the Mitigation Controller (MitC) has been developed using an integrative approach of non-linear optimization techniques, probabilistic Monte Carlo simulation, and preference function modeling.

**Findings (limit 100 words)**

Compared to the conventional way of mitigating project delays. The developed MitC allows mitigating potential delays with the least negative consequences on several project criteria, such as cost, environmental impact, etc. The application of the model to the demonstrative case study shows the ability of the model to significantly increase the probability of completing the project in the given target duration. Embedding the multi-criteria evaluation in the optimization model ensures that other interests are also represented in finding the optimal strategy for project delays.

**Originality/value (limit 100 words)**

[1] Faculty of Civil Engineering and Geosciences, Delft University of Technology, Netherlands.
[2] Faculty of Technology, Policy, and Management, Delft University of Technology, Delft, Netherlands, Corresponding author Email: o.kammouh@tudelft.nl
[3] Faculty of Civil Engineering and Geosciences, Delft University of Technology, Netherlands.

The MitC is a state-of-the-art decision support tool that a-priori automates and optimizes the search for the best set of mitigation strategies for common purpose rather than a-posteriori evaluating the potentially sub-optimal and over-designed mitigation strategies. Combining simulation and optimization of the project network schedule to systematically change the network structure to mitigate the effects of risks and uncertainties is something that was not yet developed. The inclusion of a multi-criteria evaluation within optimization is also a novel perspective in the field. These two aspects prove this development to be a relevant addition to the scientific knowledge.

## 1. Introduction

Large Dutch infrastructure projects are nowadays mostly executed by means of DBFM (Design Build Finance Maintain) contracts. The rationale is that the one that can best control and mitigate the risk is responsible for that risk. In practice, this results in significant risks and responsibilities for the contracting party (Verweij & Meerkerk, 2021). A risk that occurs and is not adequately mitigated can cause significant delays to the project. According to contractual arrangements, these delays are penalized by the client, which can cause budget overruns for the contractor. Proper scheduling and risk management is nowadays the answer to identifying the effects of risks and uncertainties on project performance. However, this is not enough to control large and complex projects.

Making a solid and reliable schedule at the start of the project is one important aspect to do in controlling your project, but dealing with dynamics, uncertainties, and setbacks during the project is challenging (Lin et al., 2011). In the last decades, probabilistic scheduling methods have been developed where, by means of Monte Carlo (MC) simulation, uncertainties in the project are captured through probability distributions (Crandall, 1976). With these probabilistic scheduling methods, the effects of uncertain durations of activities and risks on the project duration can be evaluated. How to mitigate the risks or increase the timely completion probability of the project is currently done manually by trial & error.

Combining risk mitigation and scheduling techniques in an adaptive control mechanism can increase the feasibility of timely completion of the project and decrease penalties or other hindrances of occurring risks. This adaptive control mechanism should find an adequate strategy to accelerate the project and minimize the delay. This can be done in two ways; either by shortening the duration of activities or compacting the network by adjusting relationships between activities.

### *1.1 Original Mitigation Controller*
Recently, (Kammouh et al., 2021) developed the Mitigation Controller (MitC), a tool that combines optimization with Monte Carlo simulation in one method for probabilistic and network-based schedules. This tool is based on a fixed target duration at the start of the project. If the required probability for a target duration cannot be met in one MC iteration, the tool aims to shorten the duration of activities, which can be done by applying corrective mitigation measures to achieve the required probability level of the target duration. In each iteration, the project target duration and costs for these mitigation measures are examined, whereby uncertainties and risks are considered. This tool is thus adding the most effective set of mitigation measures that crashes the duration of activities as much as needed for the minimal cost. This is done by applying multi-objective linear optimization modeling.

The other perspective to accelerate a project is by changing the network's structure. Nowadays, network-based schedules assume that the assigned relations between activities are fixed. Nonetheless, it could happen that other types of logic sequences within a network are feasible as well or could even lead to a shorter project completion time (Ballesteros-Pérez, Elamrousy, & González-Cruz, 2019). Considering logic sequences in a network as *soft* (i.e. adaptive), gives the project team more flexibility and freedom to mitigate delays (Jaskowski & Sobotka, 2012). This is not yet implemented in the original Mitigation Controller.

In addition, the original MitC only examines cost as a criterion for the best mitigation strategy (i.e. set of mitigation measures) to minimize the delay. However, as large infrastructure projects are mostly commissioned by public clients, it is of interest to not only consider cost as a criteria to find the optimal mitigation strategy, but to also consider other type of criteria for adjustments in the schedule. The effects of a mitigation strategy can be considered using a multi criteria evaluation within the optimization to improve the goal-oriented behavior of all project stakeholders in the model.

### 1.2 Soft links in a network

Little research is done on the adjustments of fixed relations in original network scheduling methods. The first distinction that should be made is whether relations cannot be adjusted, so-called fixed relations, or can be adjusted, which can be seen as *soft* relations. Relations can be fixed due to their logical characteristic or resource dependency (Khoramshahi & Ruwanpura, 2011). Activities with a soft relation are those that can be related to each other in multiple ways, which makes the relation soft (Tamimi & Diekmann, 1988). Martens & Vanhoucke (2019) describe the process of adjusting precedence relationships from finish to start to a (partially) parallel relationship as 'fast-tracking'. With fast-tracking, the network is compacted to shorten the total project duration. The model of Tamimi & Diekmann (1988) was the first scientific study on adaptive network restructuring, called SOFTCPM, and deals with logical sequences in a CPM based network that are convertible. This model can update the soft logics in the network when an unexpected event occurs, which can cause delays in the planned project duration. However, this model cannot deal with activities that partly overlap, does not classify any other logical relations as soft logic, and is entirely deterministic. The SERSI model works further on the work of Tamimi & Diekmann (1988) and distinguishes three types of soft logics: OR, EXCLUSIVE-OR and SOFT sequences. OR activities can be executed in parallel or in series, EXCLUSIVE-OR relations can be reversed, and SOFT relations can be executed simultaneously or reversed. The aim of the SERSI model is to minimize the impact of changes during the project on the initial schedule (El-Sersy, 1992)

PROSOFT (Probabilistic network scheduling model considering soft logics) is a model that does consider probabilistic network schedules. Wang (2005) does not only apply uncertainties to the activity durations, but also to the relations between activities. PROSOFT uses the same classifications of soft logics as used in the research of El-Sersy (1992). However, the model of Wang (2005) does not aim for the optimal mitigation strategy by adjusting the network. It only aims to incorporate uncertainties within the relations between activities in a project.

A general conclusion that can be drawn is that most models do not consider uncertainties and risks, and the cost is the only consequence of schedule adjustments that are considered. Besides, the current network restructuring models are not explicitly designed for mitigation strategies in case of project delays.

A model that adjusts the logical sequences in the network to mitigate delays does not yet exist. An adaptive computer tool that optimizes the strategy to compact the network in such a way that it minimizes delays can improve the ability of contractors to reduce major cost overruns. With this study, the aim is to develop an adaptive computer tool that can support contracting parties in large infrastructure projects. In addition, it is of interest to not only include cost as a factor in the trade-off but to also consider other types of criteria for adjustments in the schedule. The negative impact of a mitigation strategy will be considered using a multi-criteria analysis within the optimization process to improve the goal-oriented behaviour of professionals in the model.

### 1.3 Contributions and paper structure

Concluding from the limitations of the current knowledge on network restructuring, a tool that can adjust soft logical sequences to mitigate delays in a project would be of interest to improve project control and the reliability of project performance. Combining the systematic approach for mitigation strategies with a Monte Carlo simulation makes the approach adequate for accounting for the uncertainties. This study aims to combine the algorithm and rationale of the MitC with the scientific gap of incorporating soft logics in network schedules as a mitigation strategy. Optimization and multi-criteria analysis are combined with a Monte Carlo simulation to obtain the most effective mitigation strategy in case of project delays and considering multiple project criteria as negative consequences of adjustments to the schedule. The structure of the paper is structured as follows. The following section elaborates on the considerations and assumptions regarding the model's logic. Section 3 explains how the MC simulation and the optimization problem are set up. Section 4, 'demonstrative example', provides insights into the application of the model introduced here, in a real-time infrastructure project in the Netherlands. The paper closes with a discussion on the model's limitations, conclusions, and recommendations for further development.

## 2. Modelling assumptions and considerations

This section explains the approach and working mechanisms of the model as well as the assumptions that are made.

### *2.1 Uncertainties and risks*

In deterministic scheduling methods, both the structure of the network and the duration of the activities are fixed, leading to one deterministic outcome of the network and its duration (Agyei, 2015). Typically, schedule changes due to, for instance, delays are manually incorporated into the schedule through a trial & error process. In probabilistic schedules, on the other hand, the durations of activities vary based on a probability distribution.

To capture uncertainties in the input parameters for activity durations ($d_i$), mitigating capacities ($m_j$), and risk durations ($d_r$), the Beta-PERT distribution is used, which is the commonly used probability distribution for modeling uncertainties in project scheduling (Ranasinghe, 1994). The Beta-PERT distribution is a commonly used distribution for probabilistic scheduling. The collection of data for this distribution is considered as advantageous and the distribution relatively easy to use in practice. Respectively, the probability density functions of these parameters are $f(d_i, a_i, m_i, b_i)$, $f(m_j, a_j, m_j, b_j)$ and $f(d_r, a_r, m_r, b_r)$, where *a, m, b* are the optimistic, most likely, and pessimistic values, respectively. The expected values in the Beta-PERT distribution are calculated as follows:

$$F[X] = \frac{a+4m+b}{6} \quad (1)$$

Whether a risk occurs in a project is considered as a binary random variable X which takes the value 1 with a probability $p_e$ when the risk occurs, and 0 for the contrary:

$$f(X; p_e) = \begin{cases} q_e = 1 - p_e & \text{if } X = 0 \\ p_e & \text{if } X = 1 \end{cases} \quad (2)$$

The duration of the risk is thus determined by the duration of the risk $d_r$ given that the risk occurs (X=1).

### *2.2 Mitigation measures*

An essential step in the model is how and if relationships in the network are soft and can be adjusted, which is defined as model input. The model can then pick the best adjustment(s) in the network structure to minimize the project delay. This restructuring process will be going through in each run in the simulation until the fixed number of iterations (*n*) is completed. If no delay is identified in an iteration, no adjustments will be made, and the network structure will not be changed.

Applying adjustments to the network structure is the main approach to mitigate delays. From these possible adjustments, mitigating capacities are calculated. The mitigating capacity of a relationship adjustment relies not only on the type of adjustment but also on the duration of the predecessor and successor in the relationship. The mitigating capacity can be interpreted as the number of overlapping days between the two activities which cannot exceed the duration of the successor activity (table 1).

*Table 1 Mitigating capacities*

| Mitigating relationship adjustment | Adjustment percentage | Mitigating capacity |
|---|---|---|
| No adjustment possible | - | 0 |
| Overlap activities with …% completion lag | 0-100% | Min(…% * duration predecessor, duration successor) |
| Release relationship | 100% | Min(duration predecessor, duration successor) |

The mitigating capacity of each possible adjustment in the schedule depends on the adjustment percentage, duration of the predecessor, and the duration of the successor (table 1). The project team can indicate a register of all soft links in the network and to what extent the relationship is soft, which can be transformed to the mitigating capacities of each soft link, based on the calculations in table 1.

### 2.3 Multicriteria evaluation

Considering multiple consequences of network adjustments by incorporating a multi-criteria analysis in an optimization problem combined with project delay mitigation is not studied so far, although it can increase the goal-oriented behavior in the model. The analysed research on incorporating soft logics in scheduling methods considers only cost as the negative effect of applying changes to the network structure. However, the ability to assign multiple criteria as consequences of the adjustment to the network structure asks for new development.

The model user should indicate the mitigation effect of each soft link in the network on multiple predefined criteria. These effects can be normalized to a preference rate from 0 (worst) to 100 (best). The worst on the normalized scale is when the combination of all mitigation measures together is applied. The best is then when no mitigation measures at all are applied. On this scale from 0 - 100, all mitigation measures and combinations of mitigation measures are rated. Hence, an aggregated preference score can be determined for each soft link (figure 2). As a result of the different units of the criteria, the aggregated preferences of the mitigation measures may not simply be added when a combination of mitigation measures is used. Therefore, the aggregated preference of all possible combinations of applied soft links should be evaluated separately, using the proper aggregation principles of the theory of preference function modelling (Barzilai., 2022)

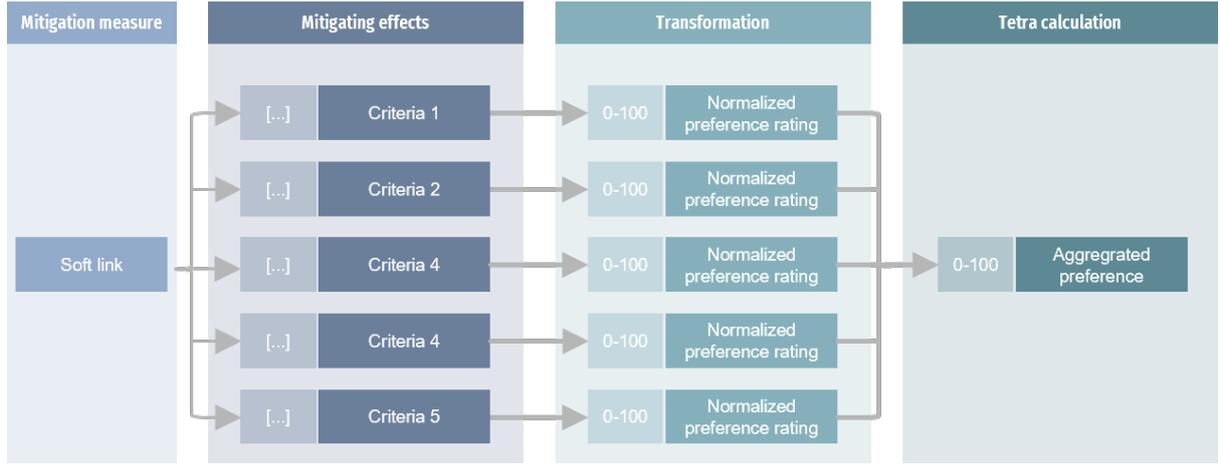

*Figure 2 Multi-Criteria Evaluation*

The optimal set of mitigation measures is the one that solves the delay as much as possible with the highest possible aggregated preference, which implies the minimal negative side effects. The objective of the model is thus to minimize the negative side effects by mitigating the delay as much as feasible.

*2.4 Relations between activities, risks and mitigation measures*

To understand how relationship adjustments can mitigate project delays, it should be clear which mitigation measures $j \in [1, 2, ..., J]$ affect certain paths *(k)* in the network. Here, the model differs from the original Mitigation Controller (MitC) as the mitigation strategies of the MitC affect specific activities while the mitigation strategies of the model presented here affect sequences between activities in the network. Therefore, from the user input on soft links in the network, the matrix $\mathbf{R_{KxJ}}$ is constructed (Eq. 3.1). $\mathbf{R_{KxJ}}$ (Eq. 3) is a matrix whose coefficients $[R_{kj}] \in \{0,1\}$ indicate if the soft link of mitigation measure *j* is on path *k*. The coefficient $R_{kj}$ takes the value 1 if the soft link of mitigation measure *j* is on path *k,* and 0 otherwise.

$$\mathbf{R_{KxJ}} = \begin{bmatrix} R_{11} & \cdots & R_{1j} \\ \vdots & \ddots & \vdots \\ R_{k1} & \cdots & R_{kj} \end{bmatrix} \quad (3)$$

## 3. Implementation of the MitC 2.0

The model aims to mitigate the delay as much as possible by adjusting relations in the network while maintaining minimal negative effects. The conceptual algorithm in figure 1 entails the modelling process for every run in the Monte Carlo simulation to come to the most effective restructured network.

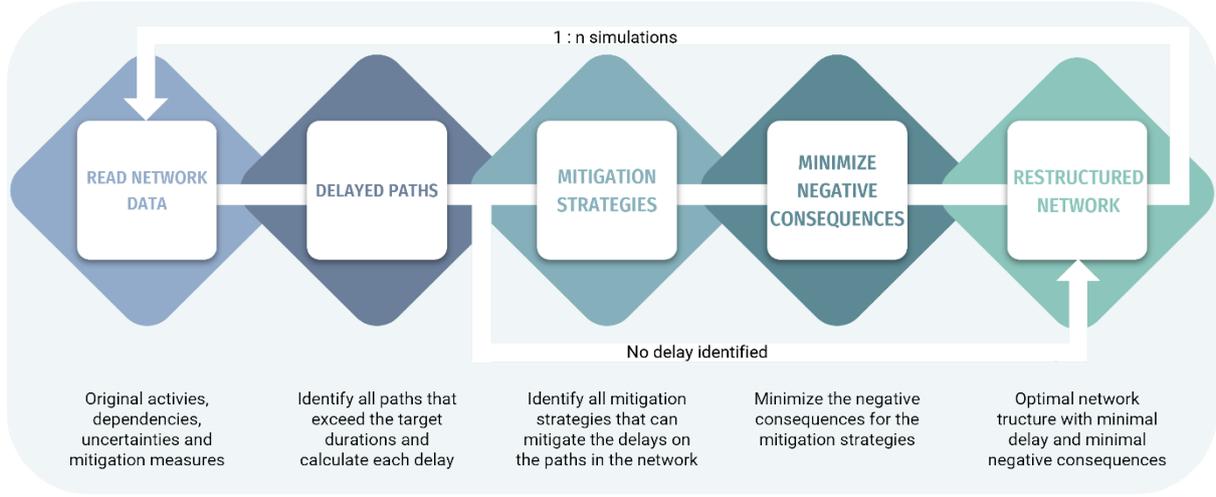

*Figure 1 Restructuring process in the simulation*

### *3.1 Mathematical formulation of the optimization problem*

The model first constructs a network that represents the project's schedule. Then, the current/non-mitigated project's completion probability is calculated to see if a mitigation process is needed. If the project completion probability is lower than the minimum accepted threshold, mitigation is then needed. In this case, the model finds the optimal mitigation strategy to complete the project within the target duration. To accomplish this, the rationale of the algorithm developed in the MitC. is used (Kammouh, et al., 2022)

A schedule network consists of a set of activities $i \in [1, 2, …, I]$, which, due to links between these activities, together form a network with a set of paths $k \in [1, 2, …, K]$ that connect the start to the finish of the project. Delays are identified when the duration of path $k$ ($D_k^0$) exceeds the planned duration ($T^{pl}$).

In each iteration, the most optimal set of mitigating adjustments to the network is found. The objective of the model entails the minimization of the negative effects (aggregated preference) of the adjusted relationships while reducing the project delay as much as feasible. The objective function is formulated below:

$$\min_{X,u} -\tilde{P}(\mathbf{X_J}) + 10^6 * u \quad (4)$$

Subject to:

$$-\mathbf{R_{KxJ}X_J} \leq T^{pl} - D_k^0 + u \quad (5)$$

where $\tilde{P}(\mathbf{X_J})$ is the aggregated preference of the combination of applied mitigation measures $\mathbf{X_J}$; and $u \in [0, 1, 2, …, \infty]$ is the utility variable which gives flexibility when the mitigation measures cannot solve the complete delay regarding the target completion date; The decision variable $\mathbf{X_J}$ is a column vector of integers [$J$ x 1] for all mitigation measures ($\mathbf{X_J} \in \{0,1\}$), where 1 means that the mitigation measure is used in the optimization strategy and 0 means the contrary. The negative of the corresponding aggregated preference will be minimized. Indicator $u$ is used as a utility variable to reflect the 'as much as feasible' constraint for the identified delay for all paths in the network. By adding Eq. 5 to the model, the constraint of completely solving the delay for every path is not a rigid constraint anymore as it could be the case that not the total delay can be mitigated by only adjusting the network. The delay on each path, which is the difference between the actual duration of the path $D_k^0$ and the planned duration($T^{pl}$), should be smaller than the mitigating capacity for the corresponding paths ($-R_{KxJ}X_J$). Utility variable $u$ takes an increasingly higher value if the model cannot generate a solution otherwise. However, as minimizing $u$ is expressed in the objective function, the model will always search for the smallest number of $u$ as feasible. The addition of the very large value to the objective function results in the fact that mitigating the delay is the most important term in the objective function to minimize. Subsequently, the combination of relationship adjustments that result in the least negative impact (highest aggregated preference) will be found which will reduce the delay as much as possible.

Another important implication of the model is that mitigation strategies that affect the same relationships between activities may not be chosen together in the same optimization. For example, the relation between activity A and B can be partially overlapped with a certain aggregated preference, but the same relationship can also be released for a lower aggregated preference. However, the computer may not choose both mitigation measures together in one optimization as they affect the same relationship. A constraint to prevent this from happening is based on mitigation measures with the same successor and predecessor. A **JxJ** matrix is constructed where mitigation measures with the same predecessor and successor have the value 1, and 0 otherwise. The sum of the product of this parameter *samepredsucc$_{jj}$* and the decision variable $\mathbf{X_J}$ should be smaller than or equal to 1:

$$\sum SAMEPREDSUCC_{JxJ}X_J \leq 1 \quad (6)$$

Lastly, constraints can be set on maximum allowable values per criteria, where the effect of mitigation measure $j$ on criteria $c$ multiplied by the decision variable $\mathbf{X_J}$ may not exceed the maximum allowable effect of criteria $c$ ($\mathbf{C_{CMAX}}$). Hereby, the optimization towards the target duration can be limited by the maximum allowable effect per criteria independently.

$$C_{CxJ} * X_J \leq C_{C\,MAX} \quad (7)$$

### 3.2 Network restructuring process with simulation and optimization

This section elaborates on the network restructuring process of the model in the computer with the simulation and optimization approach as described in previous sections. The code is written in Matlab version R2021a and uses the Optimization toolbox and Global Optimization toolbox.

At first, all network characteristics, mitigation characteristics, and the target duration are read. After that, the Monte Carlo simulation can run for $n$ runs. The number of runs can be indicated by the model user. In every iteration, the following steps should be taken. At first, the network should be constructed based on the original dependencies between activities. It is assumed that, in the original network, all dependencies are FS relations. The model will return this information in a network with nodes and arrows, where the nodes are the activities, and the arrows represent the predecessor's duration before the succeeding activity in the network can start. When this network is constructed, the MC simulation can start. In each run, random durations will be assigned to the activities according to the given probability distribution.

Additionally, risks will be assigned to the affected activities in the network according to their probability of occurrence. Then, the mitigating effect of each relationship adjustment will be calculated. The mitigating effects will be calculated in such a way that they can be subtracted from the affected paths. The next step is to define all paths in the network and calculate the length of each from start to finish. For every path, the delay should be calculated relative to the given target duration. The delay is 0 if the path is shorter than or equal to the target duration. When all these calculations are done, the optimization can start.

For the optimization problem, a solver-based approach is used as the optimization goal is predetermined. The objective function in the optimization is non-linear due to the separate evaluation of aggregated preference of different combinations of mitigation measures, using a software tool called Tetra for the proper mathematical aggregation (i.e., Tetra is a preference function based solver to evaluate the aggregation of different decision/design alternatives, see Barzilai, 2010 and/or see Scientific Metrics, 2022). Although all other calculations and constraints are linear, non-linear optimization is thus required. As all decision variables are integer a Genetic Algorithm suits the best for this optimization problem. As the optimization takes place in every iteration of the MC simulation, the optimization results in every run are stored and used for descriptive and statistical analysis of the results which can form the basis of conclusions and recommendations to the project manager on the mitigation strategy that should be applied. As the optimization takes place in every iteration of the MCS, the optimization results in every run are stored and used for descriptive and statistical analysis of the results which can form the basis of conclusions and recommendations to the project manager on the mitigation strategy that should be applied.

## 4. Demonstrative case study

To analyse results and validate the model, a demonstrative case study is applied to the developed model.

### *4.1 Project characteristics*

The A16 Rotterdam project is currently one of the largest ongoing infrastructure projects in execution in the Netherlands. This highway project contains a trajectory of 11 kilometres and aims to be the first energy-neutral highway in the Netherlands. The trajectory connects traffic node Tebregseplein on the A16 with the A13 at the northern part of Rotterdam. For this project, a DBFM contract is used with a maintenance period of 20 years, which implies the project organization is responsible for the Design,

Build, Finance, and Maintenance of the project. The project organization gets paid by the client based on delivered goods instead of services and should prefinance its project by external financial parties and pays back this loan when the delivered goods are finished. Therefore, the project should be prevented from delays as much as possible so that loans can be paid back as soon as possible with the payments from the client within the agreed time.

The entire A16 Rotterdam project consists of several subprojects, including a tunnel. A tunnel is often part of large infrastructure projects where DBFM contracts are applied. Therefore, using the tunnel as a case study is interesting due to its reproducibility to other DBFM projects. Within this tunnel project, it is chosen to evaluate and optimize the schedule for the construction pits that need to be constructed as part of the tunnel. The entire tunnel project consists of 36 construction pits. Analysing the optimization of scheduling construction pits cannot only be reproduced to other tunnel projects, but also within the Rottemerentunnel zooming in on analysing one construction pit can be reproduced to other construction pits within the Rottemerentunnel.

The data that is used for this analysis is the original schedule for all individual construction pits that was set up at the start of the entire A16 Rotterdam project. For the construction pits that are already in execution, changes have already been made to this schedule due to occurring risks and unforeseen events. For the specific case study, one individual construction pit, it will be analysed whether adjustments in sequences in the network could have prevented delays from a historical point of view. Summarized, the following data is gathered from the project organization:

- Export of original schedule of the individual construction pit
    - Network characteristics with only FS relations between activities
    - Three-point estimates for activity durations
- Risk register
    - Three-point estimates of risk impact (delay duration due to risk)
    - Probability of occurrence of risks
- Register of mitigation measures
    - Soft links that can be compacted or released
    - Three-point estimates of mitigating capacities of each soft link
    - Mitigating effects for the 5 defined criteria of each soft link
- Costs of activities in the original FS network

The activities that are part of the construction process are listed below:

*Table I Activity register*

| Activity | | Activity duration | | | Predecessors |
|---|---|---|---|---|---|
| ID | Description | Optimistic | Most likely | Pessimistic | Activity ID |
| 1 | Start project | 0 | 0 | 0 | |
| 2 | Apply permit new gas pipe | 48 | 50 | 58 | 1 |
| 3 | Digging trench gas pipe | 18 | 20 | 22 | 2 |
| 4 | Install gas pipe | 22 | 25 | 29 | 3 |
| 5 | Closing trench gas pipe | 18 | 20 | 22 | 4 |
| 6 | Installing sheet piles 11 | 13 | 15 | 17 | 1 |
| 7 | Installing sheet piles 12 | 13 | 15 | 17 | 6 |
| 8 | Installing pile foundations | 27 | 30 | 33 | 6 7 |
| 9 | Excavation bulk to top of piles | 4 | 5 | 6 | 8 |
| 10 | Excavation between piles | 9 | 10 | 11 | 9 |
| 11 | Sludge suction | 13 | 15 | 17 | 10 |
| 12 | Crack piles and clean site | 13 | 15 | 17 | 11 |
| 13 | Measuring piles | 9 | 10 | 11 | 12 |
| 14 | Earthworks gravel | 4 | 5 | 5 | 16 17 |
| 15 | Installing reinforcement underwater concrete | 36 | 40 | 44 | 14 5 |
| 16 | Pouring underwater concrete | 1 | 1 | 2 | 14 15 |
| 17 | Harden underwater concrete | 9 | 10 | 11 | 18 |
| 18 | Emptying construction pit | 4 | 5 | 5 | 16 17 |
| 19 | Cleaning underwater concrete | 9 | 10 | 11 | 18 |
| 20 | Install construction joint | 4 | 5 | 6 | 18 19 |
| 21 | Lag 15-19 | 5 | 5 | 5 | 19 |
| 22 | Lag 16-19 | 5 | 5 | 5 | 20 |
| 23 | Part 019 – Construct inner walls | 4 | 5 | 6 | 19 20 21 22 |
| 24 | Lag 19-21 | 9 | 10 | 11 | 23 |
| 25 | Part 019 – Construct outer walls and deck | 4 | 5 | 6 | 19 23 24 |
| 26 | Part 020 – Construct inner walls | 4 | 5 | 6 | 23 |
| 27 | Part 020 – Construct outer walls and deck | 4 | 5 | 6 | 25 26 |
| 28 | Part 021 – Construct inner walls | 4 | 5 | 6 | 26 |
| 29 | Part 021 – Construct outer walls and deck | 4 | 5 | 6 | 27 28 |
| 30 | Part 022 – Construct inner walls | 4 | 5 | 6 | 28 |
| 31 | Part 022 – Construct outer walls and deck | 4 | 5 | 6 | 29 30 |
| 32 | Part 023 – Construct inner walls | 4 | 5 | 6 | 30 |
| 33 | Part 023 – Construct outer walls and deck | 4 | 5 | 6 | 31 32 |
| 34 | Part 024 – Construct inner walls | 4 | 5 | 6 | 32 |
| 35 | Part 024 – Construct outer walls and deck | 4 | 5 | 6 | 33 34 |
| 36 | Part 025 – Construct inner walls | 4 | 5 | 6 | 34 |
| 37 | Part 025 – Construct outer walls and deck | 4 | 5 | 6 | 35 36 |
| 38 | Part 026 – Construct inner walls | 4 | 5 | 6 | 36 |
| 39 | Part 026 – Construct outer walls and deck | 4 | 5 | 6 | 37 38 |
| 40 | End project | 0 | 0 | 0 | 39 |

Figure 3 represents the deterministic schedule where no risks or uncertainties are considered. There is one clear critical path (red). The project will then last for 268 days.

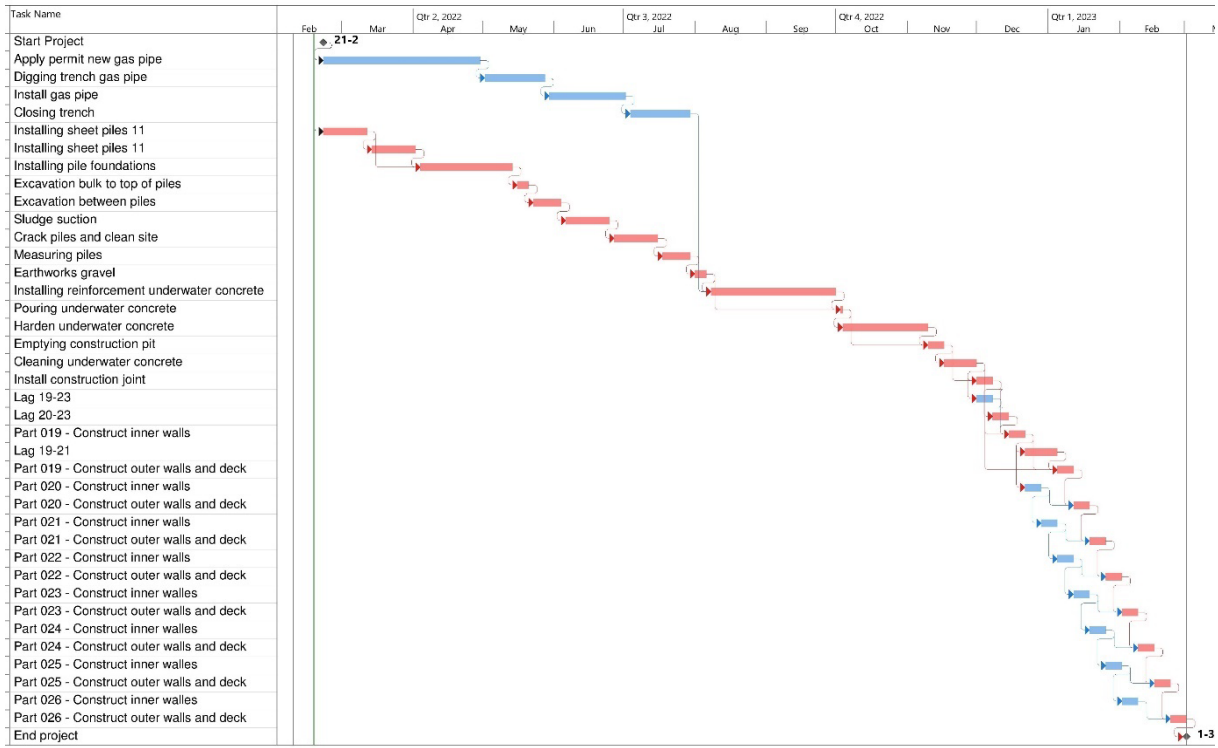

*Figure 3 Original schedule for construction pit of the Rottemerentunnel*

*Table 3 Risk register*

| Risk | | Risk duration | | | Risk events and activities | Risk probability |
|---|---|---|---|---|---|---|
| ID | Description | Optimistic | Most likely | Pessimistic | | |
| 1 | Strong wind during crane use | 0 | 1 | 5 | 11 | 0.38 |
| 2 | Capacity shortage of supplier reinforcement | 5 | 10 | 15 | 11 | 0.18 |
| 3 | Capacity shortage concrete batching plant | 2 | 5 | 10 | 12 | 0.5 |
| 4 | Not enough drivers or containers | 5 | 12 | 20 | 8 | 0.5 |
| 5 | Water from the tunnel cannot be discharged | 5 | 10 | 20 | 14 | 0.05 |
| 6 | Cracking and leakage in the construction pit | 5 | 12 | 20 | 16 | 0.18 |
| 7 | Pile foundation: installation does not comply with design | 8 | 16 | 20 | 11 | 0.5 |
| 8 | Use of sheet pile and pile scaffolding stalls | 10 | 15 | 20 | 4 | 0.38 |
| 9 | Excavation tunnel takes longer than expected | 20 | 30 | 40 | 6 | 0.38 |
| 10 | Bigger outer pile decreases production rate | 15 | 20 | 30 | 4 | 0.38 |
| 11 | Grout on pile heads should be removed | 10 | 15 | 25 | 8 | 0.38 |

| 12 | Gravel rises because of a swell in the subsoil | 5 | 10 | 20 | 11 | 0.01 |
| 13 | Enforcer "Hoogheemraadschap" quits work | 5 | 12 | 20 | 14 | 0.1 |
| 14 | Reduced quality of underwater concrete causes leakage | 5 | 10 | 15 | 15 | 0.1 |
| 15 | Obstacles in the soil causes hindrance for pile foundation | 1 | 2 | 5 | 4 | 0.03 |
| 16 | Capacity shortage concrete batching plant | 2 | 5 | 10 | 19 | 0.5 |
| 17 | Permit disapproval | 10 | 14 | 16 | 2 | 0.1 |

However, when the risk register (table 3) is included and not only the most likely impact of the risk are considered, it will become problematic to finish the project within 268 days. Soft links that can fasten the project and thus mitigate delays are defined by the project team. The criteria determined to define the negative effects of the mitigation measures are investment cost, operational cost, probability of failure, stakeholder effects, and traffic hindrance. The investment and operational costs are measured in euros. The probability of failure is defined as the amount of money needed to redo work due to failures at the construction site. As the project risks have a relatively high probability of occurrence and according to the wish of the project team, a scenario is tested to restructure the network to retrieve the highest probability of completing the project within 300 days. Some extra days relative to the deterministic project duration are thus accepted. Moreover, constraints will be considered on the total budget, penalty points, and lost vehicle hours of the chosen mitigation measures. The constraints are defined as follows:

- Total investment cost ≤ €80.000,-
- Total operational cost ≤ €60.000,-
- Total cost for failure ≤ €8.000,-
- Total received penalty points ≤ 4
- Total lost vehicle hours ≤ 15.000

*4.2 Results*

Including these constraints on the defined project criteria in the optimization is important to represent other project interests than only delivering the project within the target duration. Although completing the project within a certain target duration is the main goal of using this model, it is important to consider other interests and criteria. Due to the budgets of the project organization, the available money for additional investments, operational costs, and failures are fixed. Due to the possibility of a negative attitude towards the project due to too many effects on stakeholders, a maximum is set on four fictive penalty points for the entire mitigation strategy. Lastly, to limit the hindrance to surrounding traffic, not more than 15.000 lost vehicle hours are allowed in the chosen mitigation strategy.

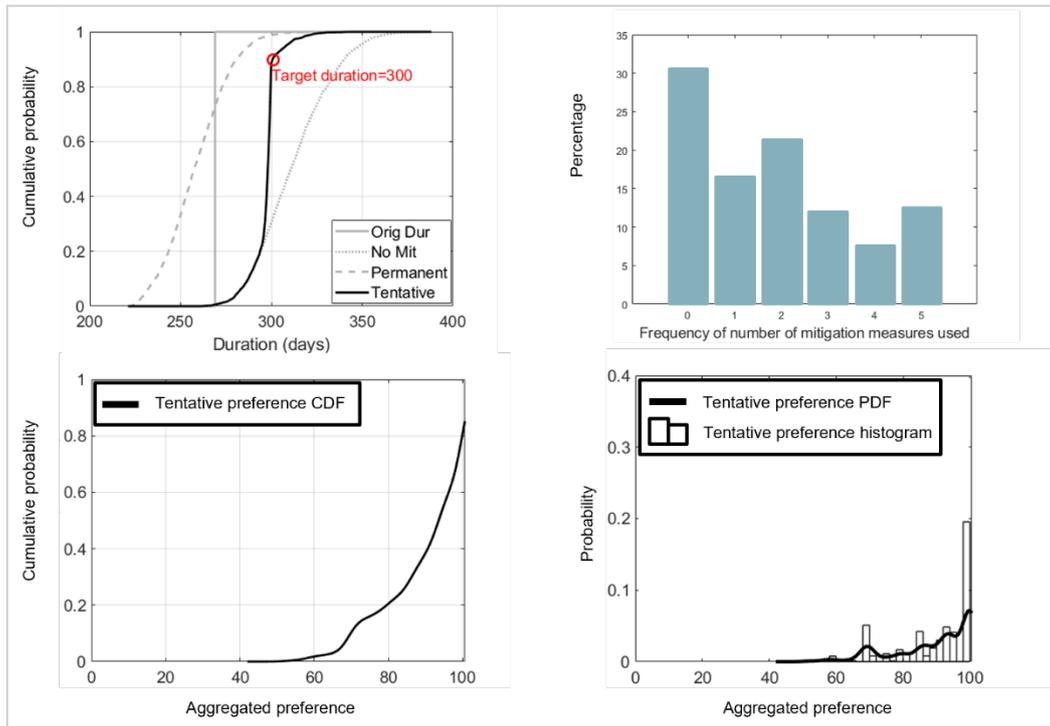

Figure 4 Model results for target duration = 300 days

The plots in figure 4 demonstrate the model results for the information given above. The probability of completing the project within 300 days is 90% against around 30% when no mitigation strategy is applied. When applying all mitigation measures, almost 100% probability can be attained. However, this mitigation strategy is not achievable due to the set constraints on the criteria in the optimization problem. Furthermore, it can be concluded that in more than 75% of the cases, 0-3 mitigation measures were applied. The cumulative distribution function and probability density function of the aggregated preference of the applied mitigation strategy demonstrate the outcome of the multi-criteria evaluation of the total negative effects of all possible mitigation strategies.

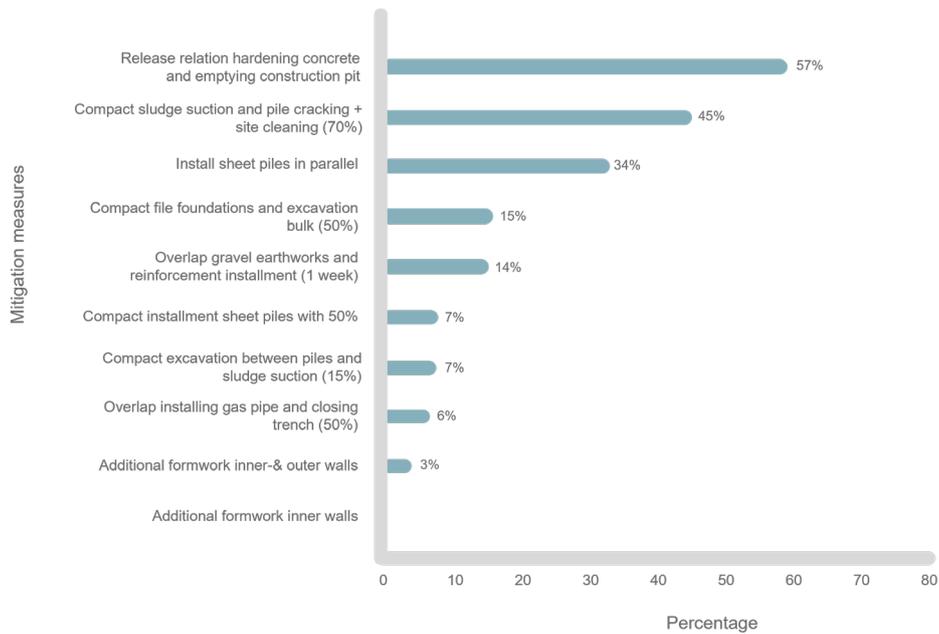

*Figure 5 Frequencies of applied mitigation measures*

Figure 5 summarizes which measures are most often chosen as part of a mitigation strategy in the optimization. Combining this information with the fact that in more than 75% of the cases 0-3 mitigation measures were applied, it would be advised to implement the first three mitigation measures as an overall strategy to increase the probability of completing the project within 300 days. By including these first three mitigation measures as a strategy, the effects on project performance can be summarized as follows:

*Table 4 Summary of the proposed mitigation strategy*

| Mitigation strategy | Investment cost | Operational cost | Failure cost | Stakeholder effect | Traffic hindrance | Aggregated preference | Probability for project completion |
|---|---|---|---|---|---|---|---|
| 3 most frequently used mitigation measures | €9.900,- | €7.020,- | €2.738,- | 2 penalty points | 2.600 lost vehicle hours | 85,25 | 70% |

The effects on the project criteria do not exceed the given boundaries. The most important conclusion that can be drawn is that the probability of completing the project within 300 days has increased from 30% to 70% by including the proposed 3 mitigation measures in the project. The consequence of this strategy is calculated to an aggregated preference of 85,25 on a scale from 0-100, where 0 is all mitigation measures at once and 100 is no mitigation measures applied. In practice, mostly 85% probability to ensure a certain project duration is required. However, as this 70% is achieved by only changing the structure of the network, this is an acceptable outcome. 70% can be seen as an acceptable probability that is achieved by only changing the structure of the network.

*4.3 Validation and practical relevance*

The members of the project organization acknowledge the added value of this model in quantifying trade-offs in the decision-making process on adjustments in a project schedule. Currently, when adjustments in the schedule are needed, and multiple options are available, the so-called 'trade-off matrices' are set up where the pros and cons of each possible adjustment are pointed out. However, these pros and cons are not quantified, and the final trade-off is based on discussion and feeling rather than on quantitative analysis and calculations. This model can effectively support the qualitative trade-off with a sound modelling process with quantified results.

Regarding the outcomes of this specific case study, the members of the project organization mentioned that they were surprised by the significant increase in the probability of completing the project within 300 days by only applying three mitigation measures. The mitigation measures that were chosen as a strategy were, according to their experience, no unusual or ineffective measures to apply. This shows that the evaluation of the defined project criteria and the calculation of mitigating capacities of each strategy on the project generate results that can be applied in real-time projects. No unexpected outcomes were returned by the model and the model can come up with a strategy that ensures a significant probability of timely completion of the project.

## 5. Discussion

The application of the model to the demonstrative case study shows the ability of the model to significantly increase the probability of completing the project in the given target duration. Embedding the multi-criteria evaluation in the optimization model ensures that other interests are also represented in finding the optimal strategy for project delays. The application of this model would especially be relevant in the initial phase of a project. Although for large and complex projects, analysis of risks and uncertainties and their effects are already common practice, it would improve the risk management to not only understand which risks are critical, but also know upfront what would be the best strategy to prevent the project from delays as much as possible.

Combining simulation and optimization of the project network schedule to systematically change the network structure to mitigate effects of risks and uncertainties is something that was not yet developed. The inclusion of a multi-criteria evaluation within optimization is also a novel perspective in the field. These two aspects prove this development to be a relevant addition to the scientific knowledge.

*5.1 Assumptions and limitations*

Although the added value of this model to the construction industry is proven, some aspects of the current mode are limited by assumptions. Firstly, it is important to mention that the model can only process original schedules that are constructed with solely FS relations. From this most conservative schedule, the model can optimize to the most effective structure of the network as a whole.

Secondly, almost only linear relations between variables are assumed. The original Mitigation Controller is completely based on linearity. As this model is an alternative perspective on the Mitigation Controller, it is chosen to retain as much coding and mathematics of the original Mitigation Controller as possible. However, the multi-criteria analysis behind the aggregated preferences of the applied mitigation measures makes the objective function in the optimization problem non-linear. This is due to the separate evaluation of each combination of mitigation measures.

Lastly, the calculations that are used in the computer program are not able to understand and translate different types of relationships in the network as scheduling software as Primavera and Microsoft Project can. Although this limitation of the current software that is being used is something to solve in the future, for now, this limitation is solved mathematically. The compacted relationship types that are used are calculated by the mitigating capacity of its application on the affected paths.

*5.2 Future work*

Aspects that are not covered in the current version of the model but are recommended to include in future development and research:

- Developing an integrated a-priori decision making tool where the extended Mitigation Controller will be integrated with the Best Fit for Common Purpose decision optimization for Construction Management ((Zhilyaev et. al., 2022). This will enable a correct mathematical modelling of the different stakeholder inputs during construction projects on-the-run, integrating multi-objective optimization, Monte-Carlo analysis and Preference Function Modelling, using the latest version of a software tool Tetra (i.e., a preference function aggregation solver, see Barzilai, 2010 and/or see Scientific Metrics, 2022).
- Dependencies or correlations between mitigation measures can exist, which are not yet included in the model.
- The mitigating capacity of a measure is now fixed to a three-point estimate with certain effects. However, these can also be functions of each other so that for each mitigation measure curves can be determined that represent the relation between an increased mitigating capacity and its effect on the project criteria.
- As this model is constructed as an alternative perspective on the Mitigation Controller it would be of interest to further research whether the original Mitigation Controller and this development can be combined into one risk mitigation computer tool. Activity crashing and network compacting can then be combined in the most optimal mitigation strategy to mitigate delays.

**Acknowledgments**

The authors thank De Groene Boog B.V.(a Dutch special contractor company) for providing the demonstrator case and for giving the possibility to verify and validate the extended MitC. Moreover, the authors would also like to thank Han Pekelharing (Primaned B.V.) and Ruud Binnekamp/ Dimtriy Zhilyaev from TU Delft for their valuable feedback.